\documentclass[%
 reprint,
 amsmath,amssymb,
 aps,
]{revtex4-2}

\usepackage{graphicx}
\usepackage{dcolumn}
\usepackage{bm}
\usepackage{tabularx,colortbl}
\usepackage[dvipsnames]{xcolor}
\definecolor{mygray}{gray}{0.8}
\usepackage{hyperref}

\begin{document}

\preprint{APS/123-QED}

\title{Dynamics of a ring resonator under concurrent index and loss modulation}

\author{Awanish Pandey}
\affiliation{%
Photonics Research Group, UGent-imec, Belgium, 9000\\ 
}%

\date{\today}

\begin{abstract}
We report a complete steady-state solution to the time rate equations governing the dynamics of a sinusoidally driven ring modulator. Compared to previous works that offer comprehensive solutions for pure index (phase) or loss (amplitude) modulation, this work provides an exact solution with their simultaneous modulation. Furthermore, we discuss the optimum modulation parameters for efficient microwave-to-optical conversion under such concurrent modulation. As an application, we report selective downward frequency conversion from the ring modulator which is not possible from pure loss or index modulation. The work allows for more accurate modeling of the ring modulator. It provides an insight to explore the interplay between phase and amplitude modulation in MRMs for unique applications and fully understand the limitations of such devices.
\end{abstract}

\maketitle
Micro ring modulator (MRM) is a vital component in silicon photonics that offers a compact footprint, large bandwidth, moderate driving voltage, and CMOS compatible fabrication \cite{mrm_review1,mrm_review2,mrm_review3}. 
Recently, there has been a growing interest to utilize MRMs for specific purposes like realizing frequency combs \cite{comb1,comb2}, wavelength scale optical isolators \cite{pandey_acs,popovic_ol}, in spectroscopy \cite{dutt}, quantum communication \cite{rf_to_optical} as well as to explore fundamental physical principles such as direction-dependent Rabi splitting \cite{Rabi}. It has also attracted a considerable amount of theoretical investigations to enable its accurate analysis and optimization of the modulation parameters for particular tasks \cite{mrm_theory1}. Typically, in MRMs, the optical carrier transmission is altered via a refractive index change of silicon by applying an external microwave signal. Depending upon whether the real or the imaginary part of the refractive index is changed, either the resonance frequency (index modulation) or the linewidth (loss modulation) of the micro ring is modified. Though most applications rely on the modulation of the real part, loss modulation accompanied by index modulation results in interesting phenomenon like non-reciprocal amplifiers \cite{non_rec_gain}, frequency-conversion \cite{freq_conv}, and light-trapping \cite{light_trap}.

  In most theoretical efforts, a pure index modulation \cite{mrm_theory1}, or pure loss modulation \cite{sacher} is assumed. At the same time, other parameters are supposed to remain fixed during the entire modulation cycle. Recently, an exact solution for the time rate equation governing the dynamics of index modulation-based MRMs were reported \cite{mrm_theory1,mrm_theory2}. It aimed to optimize the MRM response for switching and microwave-to-optical conversion applications. However, for practical carrier depletion (or injection) based silicon MRMs operating on the plasma-dispersion effect, the index modulation is always accompanied by an associated loss modulation following the Kramers-Kronig relationship. Hence, even though a pure index or loss modulation inspection provides a significant insight to design, understand the functioning, and limitations of MRMs, they are not complete. The few studies that attempt to describe MRM dynamics in the presence of both modulation types introduce approximations, e.g. assuming a linear variation of the resonance frequency and loss as a function of the applied microwave signal strength. Therefore, they are not suitable when the strength of the microwave signal is large \cite{mrm_theory3,mrm_theory4}.

This letter develops a complete steady-state solution of the MRM dynamics for simultaneous index and loss modulation without any approximation. We report the influence of the concurrent modulation on the sideband gain and the optimum condition for efficient first-order sideband conversion as a function of the optical carrier placement. In addition, we also study the variation in the critical coupling condition of the micro ring with the modulation. Furthermore, we report selective and efficient down-conversion of the frequency with zero transmission of the up-converted frequency components as an application. Such type of frequency conversion finds use in numerous functions like uni-directional inter-band modulation \cite{rakich_interband}, frequency shifting \cite{frequency_shifting,frequency_shifting1}, signal processing \cite{signal_processing}, and is typically attained by acousto-optic \cite{bound_state,loncar_ssb1,loncar_ssb2} or electro-optic modulators in specific configurations \cite{zimmerman,pandey_oe}.

\begin{figure}[htbp]
   \centering
    \includegraphics[width=70mm]{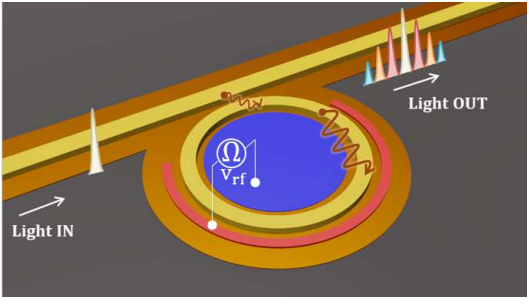}
    \caption{Schematic of the micro ring modulator showing the single frequency optical input and the series of sidebands as the output.}\label{fig:schematic}
\end{figure}

The schematic of the MRM considered in this work is shown in Fig. \ref{fig:schematic}. It consists of a bus waveguide with optical input and output port evanescently coupled to a ring modulator. The time-rate eqn. for the optical-field amplitude ($\alpha(t)$) inside the MRM, according to the coupled-mode theory \cite{cmt}, is given by:
 

\begin{equation}
\begin{split}
\label{eqn:cmt1}
  \frac{d\alpha(t)}{dt} = \iota\left(\omega_0(t) + \iota\gamma(t)\right)\alpha(t)  - \iota \sqrt{2\gamma_c} s_{in}
\end{split}
\end{equation}

and the optical output is obtained using the eqn.:
\begin{equation}
\label{eqn:cmt2}
s_{out} = s_{in} - \iota\sqrt{2\gamma_c}\alpha(t)
\end{equation}

In the above eqns., $\omega_0$ is the ring resonance frequency, $\gamma_c$ is the bus-ring coupling bandwidth, $\gamma = \gamma_l + \gamma_c$ where $\gamma_l$ is the loss bandwidth of the resonator, and $\frac{\omega_0}{2\gamma}$ is the resonance Q-factor. The optical input $s_{in} = Ae^{\iota\omega_l t}$ where $A$ is the amplitude and $\omega_l$ is the optical frequency. Under modulation from a periodic microwave signal $v_{rf}(t) = v_{rf}\cos{\Omega t}/2$, both the resonance frequency ($\omega_0$) as well as the shape of the modulator's optical transmission ($\gamma_l$) will vary as a function of time according to $\omega_0(t) = \omega_0 + \delta\omega_m(t)$ and $\gamma_l = \gamma_l + \delta\gamma_l(t)$ with $\delta\omega_m(t) = \frac{\delta\omega_m}{2}\cos(\Omega t)$ and $\delta\gamma_l(t) = \frac{\delta\gamma_l}{2} \cos(\Omega t)$. $\delta\omega_m$ and $\delta\gamma_l$ are the peak-to-peak index modulation and loss modulation amplitude respectively. The modulation amplitudes are related to the applied microwave signal as $\delta\omega_m = \frac{\partial\omega_0}{\partial v_{rf}}v_{rf}$ and $\delta\gamma_l = \frac{\partial\gamma_l}{\partial v_{rf}}v_{rf}$ with $\frac{\partial\omega_0}{\partial v_{rf}}$ and $\frac{\partial\gamma_l}{\partial v_{rf}}$ being the change in resonance frequency and linewidth of the ring resonance per volt of the applied microwave signal. The differential eqn. \ref{eqn:cmt1} is solved under the gauge transformation $\beta(t) = \alpha(t)e^{-\iota\omega_l t - \iota \left[\int_{0}^{t} \left(\delta\omega_m(t') + \iota\delta\gamma_l(t')\right)dt'\right]}$.


The transformation defines a frame where the ring is unmodulated and the optical energy inside the ring is proportional to $|\beta(t)|^2$. Due to the presence of loss modulation, $|\beta(t)|^2 \neq |\alpha(t)|^2$ in the new frame. $\beta(t)$ is related to $\alpha(t)$ by a canonical transformation invariant under the translation $t \rightarrow t + 2n\pi$. Such variable transformations,with $\delta\gamma_l = 0$ implying $|\beta(t)|^2$ = $|\alpha(t)|^2$, have been previously applied to study dynamic isolation in temporally modulated coupled harmonic oscillators \cite{salerno} and periodically driven quantum systems such as a particle moving in a modulated harmonic trap \cite{goldman}. The time evolution of the variable $\beta(t)$, using eqn. \ref{eqn:cmt1} and applying Jacobi-Anger expansion, takes the form:



\begin{equation}
\begin{split}
\label{eqn:dbeta2}
\frac{d\beta(t)}{dt} = \left(-\iota\Delta\omega - \gamma \right)\beta(t) - \iota\sqrt{2\gamma_c}A\sum_{n=-\infty}^{\infty}\sum_{k=-\infty}^{\infty} -1^n \\ J_n\left(\frac{\delta\omega_m}{2\Omega}\right) \left(\frac{1}{\iota}\right)^k I_k\left(\frac{\delta\gamma_l}{2\Omega}\right)e^{\iota(k+n)\Omega t}
\end{split}
\end{equation}

where $\Delta\omega = \omega_l - \omega_0$ is the detuning between the resonance frequency and the input optical carrier. $J_a(b)$ is the Bessel function of the first kind with an argument $b$ and order $a$, and $I_a(b)$ is the modified Bessel function of the first kind. In analogy to the eqn. \ref{eqn:cmt1}, eqn. \ref{eqn:dbeta2} represents a passive cavity with an un-modulated resonance frequency of $\Delta\omega$, a constant total decay rate of $\gamma$, and driven by an infinite number of sources placed at $(k+n)\Omega$ with their amplitudes scaled by factor $(-1)^n \left(\frac{1}{\iota}\right)^k A J_n\left(\frac{\delta\omega_m}{2\Omega}\right)I_k\left(\frac{\delta\gamma_l}{2\Omega}\right)$. To solve eqn. \ref{eqn:dbeta2}, a Fourier series decomposition of $\beta(t) = \sum_{p=-\infty}^{\infty}\beta_p e^{\iota p \Omega t}$ is performed and using eqn. \ref{eqn:dbeta2} to determine the value of $\beta_p$, we get:

\begin{equation}
\begin{split}
\label{eqn:beta_series1}
\beta_p = - \iota\sqrt{2\gamma_c}A\sum_{n=-\infty}^{\infty}\sum_{k=-\infty}^{\infty}-1^n J_n\left(\frac{\delta\omega_m}{2\Omega}\right)\\ \left(\frac{1}{\iota}\right)^k I_k\left(\frac{\delta\gamma_l}{2\Omega}\right) \frac{1}{\iota\Delta\omega +\iota (n+k) \Omega +  \gamma}  \iff n+k = p
\end{split}
\end{equation}







Applying $\alpha(t) = \beta(t) e^{+\iota\omega_l t + \iota \left[\int_{0}^{t} \left(\delta\omega_m(t') + \iota\delta\gamma_l(t')\right)dt'\right]}$ to return to the original variable $\alpha(t)$, the optical field amplitude in the ring will be:



\normalsize
\begin{equation}
\begin{split}
\label{eqn:alpha_series1}
\alpha(t) = \sum_{q=-\infty}^{\infty}\beta_qe^{\iota(\omega_l + q\Omega)t} \times \sum_{m=-\infty}^{\infty}\sum_{q=-\infty}^{\infty}J_m\left(\frac{\delta\omega_m}{2\Omega}\right)\left(\frac{1}{\iota}\right)^q \\ I_q\left(\frac{\delta\gamma_l}{2\Omega}\right)e^{\iota (m-q)\Omega t}
\end{split}
\end{equation}

Finally, the output of the MRM is determined using eqn. \ref{eqn:cmt2} to be:



\normalsize
\begin{equation}
\begin{split}
\label{eqn:master}
\frac{s_{out}(w\Omega + \omega_l)}{s_{in}(\omega_l)} = 1-  2\gamma_c\sum_{m=-\infty}^{\infty}\sum_{p=-\infty}^{\infty}\sum_{n=-\infty}^{\infty} \sum_{k=-\infty}^{\infty} [  (-1)^{n}\\ \left(\frac{1}{\iota}\right)^{k+p}J_m\left(\frac{\delta\omega_m}{2\Omega}\right) J_n\left(\frac{\delta\omega_m}{2\Omega}\right) I_p\left(\frac{\delta\omega_m}{2\Omega}\right)\\ I_k\left(\frac{\delta\omega_m}{2\Omega}\right) \frac{1}{\iota\Delta\omega +\iota (n+k) \Omega +  \gamma}e^{\iota(n+m+k-p)\Omega t}]
\end{split}
\end{equation}

\begin{figure}[htbp]
   \centering
    \includegraphics[width=\linewidth]{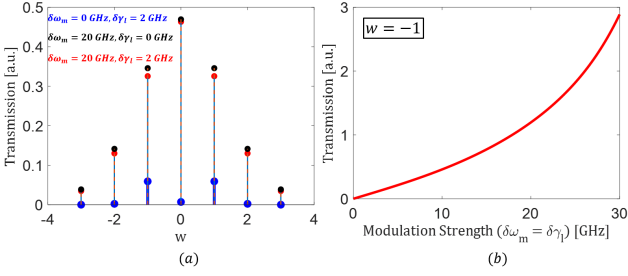}
    \caption{(a) Strength of sidebands and optical carrier with different modulation ampltitude combination, and (b) first order sideband transmission as a function of increasing modulation amplitude.}\label{fig:first_order}
\end{figure}

Eqn. \ref{eqn:master} represents the transmission of MRM with optical frequency components at $w\Omega + \omega_l$ where $w = n+m+k-p$. It is a complete solution to the time rate eqn. of the MRM without any approximations to study MRM response as long as eqns. \ref{eqn:cmt1} and, \ref{eqn:cmt2} are valid to describe its behavior. It enables the treatment of pure index modulation by setting $\delta\gamma_l = 0$, pure loss modulation by setting $\delta\omega_m = 0$, and when both index and loss are modulated together. The transmission of the optical carrier is obtained when $w = 0$ while the transmission of the sidebands is obtained when it is non-zero. For example, the transmission of third-order sideband is obtained when $w = \pm 3$, $+$ represents upper order and $-$ represents lower order. Using eqn. \ref{eqn:master}, the transmission of the optical carrier and first three orders of sideband is shown in Fig. \ref{fig:first_order} for $\gamma_l = \gamma_c = $ 3 GHz, $\Omega = $ 6 GHz, $\Delta\omega = 0$, and different modulation amplitude ($\delta\omega_m, \delta\gamma_l$). The selection of the parameters in this work is based on our recent experimental demonstration of a frequency-comb as well as an optical isolator using ring modulators on an all-silicon platform \cite{group_four,pandey_acs}. Eqn. \ref{eqn:master}, however, is general and can be used to explore a wide range of parameters depending upon the application.

In the unmodulated ring $(\delta\omega_m = \delta\gamma_l = 0)$, the carrier transmission is zero at the resonance frequency ($\Delta\omega = 0, w = 0$) as the ring is critically coupled ($\gamma_c = \gamma_l$). In contrast, in the presence of modulation, the strength of the resonance is non-zero. Under pure index and loss modulation ($\delta\omega_m = 20$ GHz), the carrier transmission is 47 $\%$, and 0.04 $\%$ respectively while for the concurrent modulation, it is 44 $\%$ as shown in Fig. \ref{fig:first_order}a. We attribute the change in the resonance extinction to the modification of the coupling condition in the presence of modulation. Towards this end, terms oscillating at $\omega_l$ in eqn. \ref{eqn:master} are collected (ignoring higher-order terms in $J_n\left(\frac{\delta\omega_m}{2\Omega}\right)$ and $I_n\left(\frac{\delta\gamma_l}{2\Omega}\right)$ for n $\neq$ 0) and are equated to zero for maximum resonance strength. The critical coupling condition, time-integrated over one modulation cycle, is then given by eqn. \ref{eqn:critical_coupling}. As evident, the critical coupling condition is modified in the presence of modulation and hence, the carrier transmission will not be null at resonance even though $\gamma_c = \gamma_l$. Under no modulation, eqn. \ref{eqn:critical_coupling} leads to the well known critical coupling condition for the passive ring resonators, $\gamma_c = \gamma_l$. 

\begin{equation}
\label{eqn:critical_coupling}
\gamma_l = \gamma_c\left[2J_0^2\left(\frac{\delta\omega_m}{2\Omega}\right)I_0^2\left(\frac{\delta\gamma_l}{2\Omega}\right) -1\right]
\end{equation}


The transmission $-1$ order sideband is shown in Fig. \ref{fig:first_order}b with varying modulation strength ($\delta\omega_m = \delta\gamma_l$). It increases with the modulation amplitude, and for sufficiently high values of $\delta\gamma_l$, it can even become larger than unity. This type of amplification is possible when the time-averaged value of $\gamma_l(t)$ becomes negative i.e. the ring modulator acts as an amplifier. Though we have discussed the transmission at $\Delta\omega = 0$, which corresponds to the maximum resonance enhancement of the carrier, the sideband conversion efficiency is also a vital parameter when the MRM is utilized as a 'mixer' for applications like microwave photonic receiver \cite{mpr}.  

\begin{figure}
   \centering
    \includegraphics[width= \linewidth]{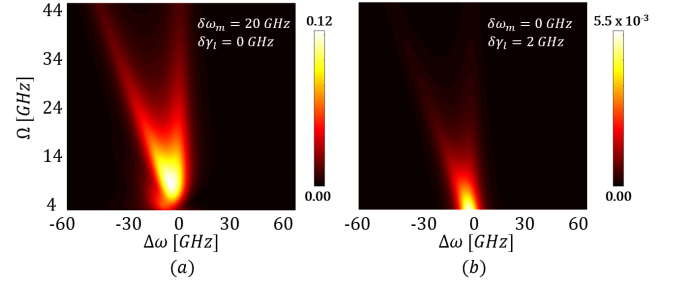}
    \caption{First order sideband strength as a function of modulation frequency and resonance detuning (a) index, and (b) loss modulation.}\label{fig:contour}
\end{figure}

To ascertain the optimum value of $\Delta\omega$ for maximum sideband strength, we plot its variation as a function of $\Delta\omega$ and $\Omega$ in Fig. \ref{fig:contour} for index and loss modulation. We observe two distinct features in Fig. \ref{fig:contour}a and \ref{fig:contour}b; the sideband strength is maximum when $\Omega < 2\gamma$, and it subsequently breaks into two paths with lower sideband strength as $\Omega$ increases. We have also checked it for concurrent modulation and found similar behavior. To quantify it, we collect the terms oscillating at $\omega_l + \Omega$ and neglect second (and higher) order terms in $J_n(\frac{\delta\omega_m}{2\Omega})$ and $\frac{\delta\gamma_l}{2\Omega}$. The transmission of the first-order sideband is then given by:


\begin{equation}
\begin{split}
\label{eqn:first_order_g}
\frac{s_{out}(\omega_l + \Omega)}{s_{in}(\omega_l)} = 2\gamma_c [J_0\left(\frac{\delta\omega_m}{2\Omega}\right)J_1\left(\frac{\delta\omega_m}{2\Omega}\right) +  \\\frac{1}{\iota}I_0\left(\frac{\delta\gamma_l}{2\Omega}\right)I_1\left(\frac{\delta\gamma_l}{2\Omega}\right)] \left(\frac{1}{\iota\Delta\omega + \gamma} - \frac{1}{ \iota\Delta\omega + \iota\Omega + \gamma}\right)e^{\iota\Omega t}
\end{split}
\end{equation}

The sideband gain is defined as the magnitude squared of eqn. \ref{eqn:first_order_g} and is given as $G = \mathcal{A} \times \mathcal{L}(\Delta\omega)\mathcal{L}(\Delta\omega + \Omega)$ where $\mathcal{L}(\Delta\omega) = \frac{\gamma^2}{\Delta\omega^2+\gamma^2}; \mathcal{L}(\Delta\omega + \Omega) = \frac{\gamma^2}{(\Delta\omega + \Omega)^2 + \gamma^2}$ are the Lorentzian functions that dictate the resonance enhancement of the optical carrier ($\mathcal{L}(\Delta\omega)$) and the first-order sideband ($\mathcal{L}(\Delta\omega + \Omega)$) with individual maximum values at $\Delta\omega = 0$ and $\Delta\omega = -\Omega$ respectively. These values individually correspond to optical carrier at the MRM resonance and the generated sideband at the MRM resonance respectively. The pre-factor $\mathcal{A}$ is written as $\mathcal{A} = \left(\frac{2\gamma_c\Omega}{\gamma^2}\right)^2 \left[J_1^2\left(\frac{\delta\omega_m}{2\Omega}\right)J_0^2\left(\frac{\delta\omega_m}{2\Omega}\right) + I_0^2\left(\frac{\delta\gamma_l}{2\Omega}\right)I_1^2\left(\frac{\delta\gamma_l}{2\Omega}\right)\right]$.




\begin{figure*}[htp]
   \centering
    \includegraphics[width= 140 mm]{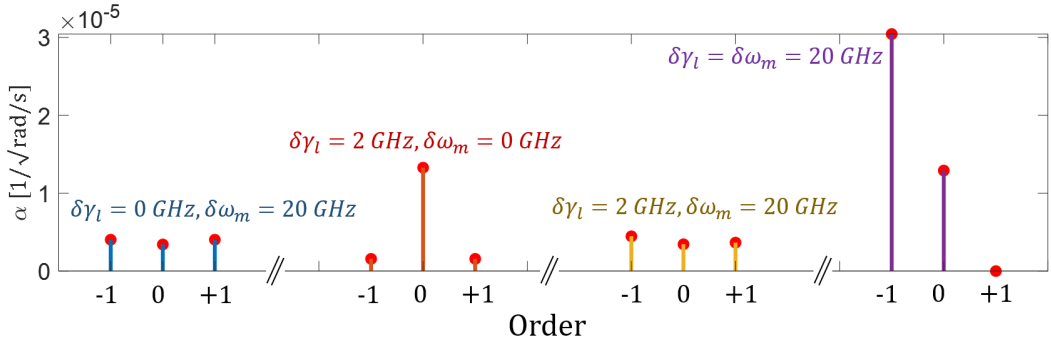}
    \caption{Amplitude of the two sidebands ($-1,+1$ order) and the optical carrier inside the ring modulator for different combinations of loss and index modulation amplitude. No $+1$ order sideband is excited in the resonator when both the modulation amplitudes are equal.}\label{fig:alpha_translation}
\end{figure*}

The optimum optical carrier placement with respect to the cavity resonance is obtained by maximizing $G$. Towards this, $\frac{dG}{d\Delta\omega} = 0$ is solved yielding three distinct values of $\Delta\omega$; $\Delta\omega = -\frac{\Omega}{2}$ when $\Omega \leq 2\gamma$, $\Delta\omega = - \frac{\Omega}{2} \pm \sqrt{\frac{\Omega^2}{2} - \gamma^2} $when $\Omega > 2\gamma$. The condition between $\Omega$ and $2\gamma$ is necessary to make sure that $\frac{d^2G}{d\Delta\omega^2} < 0$ at these particular values of $\Delta\omega$. Under these conditions, the optimum value of $G$ is given by eqns. $G\left(\Delta\omega = -\frac{\Omega}{2};\Omega\leq2\gamma\right) = \mathcal{A} \times \left(\frac{\gamma^2}{\left(\frac{\Omega}{2}\right)^2 + \gamma^2}\right)^2$, and $G\left(\Delta\omega = - \frac{\Omega}{2} \pm \sqrt{\frac{\Omega^2}{2} - \gamma^2};  \Omega>2\gamma\right) = \mathcal{A} \times \left(\frac{\gamma}{\Omega}\right)^2$. It is clear from the above analysis that in case when $\delta\gamma_{l}$ is non-negligible compared to $\Omega$, loss-modulation will have substantial influence on the sideband gain as the function $I_a(b)$ increases sharply with $b$. In particular, for $\Delta\omega = -\frac{\Omega}{2}$ and $\Omega \leq 2\gamma$, the sideband transmission can amplify to a value more than unity when $\mathcal{A}> \left(\frac{\left(\frac{\Omega}{2}\right)^2 + \gamma^2}{\gamma^2}\right)^2$.

Carrying out similar analysis for higher-order sidebands, we deduce that for $N^{th}$ order sideband, the optimal $\Delta\omega = \frac{N\Omega}{2}$ when $\Omega \leq 2\gamma$. Therefore, it is impossible to achieve maximum sideband gain for more than one sideband simultaneously as they occur at different de-tunings ($\Delta\omega$). Hence, for functions requiring uniform sideband power over many sidebands (e.g., frequency combs), a compromise needs to be made between the strength of each sideband and power uniformity.

Applying the developed theory, we report downward frequency conversion from a ring modulator with simultaneous loss and index modulation. Down-conversion of frequency implies that the ring modulator will contain only the lower sideband harmonics ($w \leq 0$ in eqn. \ref{eqn:master}). The transmission of upper sideband harmonics ($w > 0$) will be zero. It involves engineering the efficiency of the wavelength conversion of individual sidebands and selectively making it zero for the up-converted sidebands. For this purpose, we introduce a phase difference between the index and loss modulation. We take $\omega_0(t) = \omega_0 + \delta\omega_m \sin(\Omega t)$ and $\gamma_l(t) = \gamma_l + \delta\gamma_l \sin(\Omega t + \phi)$. From eqn. \ref{eqn: alpha_series1}, the field amplitude inside the ring is written as a Fourier series $\sum_{n=-\infty}^{\infty}\alpha^n e^{\iota(\omega_l + n\Omega t)}$ where $\alpha^n$ represents the complex, time-independent field amplitude of each sideband and optical carrier component inside the modulator. Plugging it in eqn. \ref{eqn:cmt1}, we get ($\delta_{n0}$ is the Kronecker delta function):

\begin{equation}
\begin{split}
\label{eqn:dispersion}
\left[\iota\Delta\omega + \iota n\Omega + \gamma_c + \gamma_l \right]\alpha^n + \left(\frac{\delta\gamma_le^{\iota\phi}}{2\iota} - \frac{\delta\omega_m}{2} \right)\alpha^{n-1} + \\
\left(\frac{-\delta\gamma_le^{-\iota\phi}}{2\iota} + \frac{\delta\omega_m}{2} \right)\alpha^{n+1} = -\iota \sqrt{2\gamma_c}A \delta_{n0}
\end{split}
\end{equation}




As evident from eqn. \ref{eqn:dispersion}, for $\phi = (4d+1)\pi/2$, the co-efficient of the harmonic components at $n-1,n+1$ will depend on the difference and sum of the index and loss modulation strength respectively. Hence, depending upon their sum and difference, the individual conversion efficiency for up- and down-converted sidebands can be modified. For pure index or loss modulation, the coefficients will have identical values. Hence, the upward ($\omega_l + \Omega$) and downward frequency ($\omega_l - \Omega$) conversion will be equally effective. Under such conditions, complete upward or downward frequency conversion is not possible. Truncating the series in eqn. \ref{eqn:dispersion} to $n = -1,0,1$, and solving the system of linear equations for $\alpha^{-1}, \alpha^0,$ and $\alpha^{+}$, we get $\alpha$ values for different harmonics as shown in Fig. \ref{fig:alpha_translation} for $\Omega = 6 GHz$, $\gamma_c = \gamma_l = 3 GHz$, and $\delta\omega = 0$. The first two sets represent values of $\alpha$ for pure index and loss modulation, respectively. The first higher and lower order sidebands have equal field amplitude inside the ring depending upon the modulation strength.

For simultaneous index and loss modulation, the efficiency of upward and downward frequency conversion can be modified. The third set in Fig. \ref{fig:alpha_translation} shows the energy amplitude for $\delta\omega_m = 20$ GHz, and $\delta\gamma_l = 2$ GHz. The amplitude for $-1$ order sideband is greater than the $+1$ order sideband. In the special case of $\delta\omega_m = \delta\gamma_l$, the co-efficient of $\alpha^-1$ collapses in the eqn. \ref{eqn:dispersion}. As seen in the fourth part of Fig. \ref{fig:alpha_translation}, it corresponds to no excitation of the $+1$ order sideband and a strong peak for the $-1$ order sideband. 

\begin{equation}
\label{eqn:alpha_1}
\frac{s_{out}(\omega_l - \Omega)}{s_{in}} = \frac{-2\gamma_c \delta\gamma_l}{[\iota(\Delta\omega-\Omega) + \gamma_c + \gamma_l][\iota\Delta\omega + \gamma_c + \gamma_l]}
\end{equation}

\begin{figure}
   \centering
    \includegraphics[width= \linewidth]{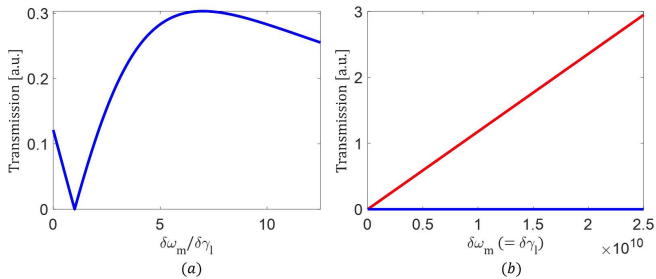}
    \caption{Transmission of (a) $+1$ order sideband with fixed $\delta\gamma_l = 2 $ GHz and varying $\delta\omega_m$ from 0 to 25 GHz, and (b) $-1$ and $+1$ order sideband with $\delta\gamma_l = \delta\omega_m$. }\label{fig:non_herm}
\end{figure}
Fig. \ref{fig:non_herm} shows the transmission of the sidebands as a function of the modulation strength. In Fig. \ref{fig:non_herm}a, $+1$ order sideband transmission is reported when the index modulation strength is varied between 0 GHz to 25 GHz keeping the loss modulation strength fixed at 2 GHz. It shows that when $\delta\omega_m = \delta\gamma_l$, the transmission of the upconverted sideband is exactly zero. At this particular condition of $\delta\omega_m = \delta\gamma_l$, transmission of $-1$ and $+1$ order sideband is given in Fig. \ref{fig:non_herm}b. The $+1$ transmission remains zero for the entire range of $\delta\omega_m$. The transmission of $-1$ order sideband exhibits a linear dependence on the modulation strength. Using CMT, a closeform equation for the output of the $-1$ order sideband is given by eqn. \ref{eqn:alpha_1}. As evident, the output is directly proportional to $\delta\gamma_l$ ($ = \delta\omega_m$). As already reported, the transmission of the sideband can experience amplification if the time-averaged (over one modulation cycle) value of $\gamma_l(t)$ becomes negative. In this particular case, this is achieved at $\delta\gamma_l = 10$ GHz.

In conclusion, a coupled mode theory-based evaluation of ring modulators in the presence of simultaneous index and loss modulation has been presented. We discussed conditions for maximum sideband enhancement and perturbation in the critical coupling parameters of the ring due to modulation. The work presented here allows tailoring the sideband strength based on the interplay between phase and amplitude modulation. We also demonstrated a frequency shifter using the modulator where only the down-converted sidebands were preserved. It was achieved by incorporating an appropriate phase difference between the index and loss modulation. The analysis presented here is for a single bus waveguide coupled to a ring modulator with moderate modulation strength. However, it can be easily extended to study a strong modulation regime where higher-order inclusion of the sideband is necessary.

\medskip
\textbf{Disclosures} The author declares no conflict of interest.

\end{document}